\documentclass[pss,fleqn]{w-art}
\usepackage{times}
\usepackage{w-thm}
\usepackage[]{epsfig,graphicx}
\begin{document}
\DOIsuffix{theDOIsuffix}
\Volume{XX}
\Issue{1}
\Month{01}
\Year{2004}
\pagespan{1}{}
\Receiveddate{2 September 2004}

\title[Scanning tunneling microscopy]{Scanning tunneling microscopy 
       of monoatomic gold chains on vicinal Si(335) surface: experimental and 
       theoretical study}

\author[M. Krawiec]{M. Krawiec\footnote{Corresponding
     author: e-mail: {\sf krawiec@kft.umcs.lublin.pl}, Phone:
+48\,81\,5376146,
     Fax: +48\,81\,5376190}} 
\author[T. Kwapinski]{T. Kwapi\'nski} 
\author[M. Jalochowski]{M. Ja\l ochowski} 
\address[]{Institute of Physics and Nanotechnology Center,
M. Curie-Sk\l{}odowska University,
 pl. M. Curie-Sk\l{}odowskiej 1, 20-031 Lublin, Poland}

\begin{abstract}
We study electronic and topographic properties of the Si(335) surface, 
containing Au wires parallel to the steps. We use scanning tunneling microscopy 
(STM) supplemented by reflection of high energy electron diffraction (RHEED) 
technique. The STM data show the space and voltage dependent oscillations of 
the distance between STM tip and the surface which can be explained within one 
band tight binding Hubbard model. We calculate the STM current using 
nonequilibrium Keldysh Green function formalism.
\end{abstract}

\maketitle


\section{Introduction}

Recently, the high-index (vicinal) surfaces have attracted much attention due 
to possibility of the creation of one dimensional structures on them 
\cite{Himpsel}. On such surfaces, deposition of small amount of metal often
results in the formation of the chain structures 
\cite{Jalochowski}-\cite{Lee}. Such one dimensional wires are important
from scientific point of view, as they should exhibit Luttinger liquid behavior 
\cite{Luttinger,Haldane}, as well as from technological one (nanoelectronics, 
quantum computing).

The examples of the one dimensional structures on vicinal surfaces are metallic 
wires on silicon, like: Au/Si(335) \cite{Crain1}-\cite{Crain}, Au/Si(557)
\cite{Jalochowski}-\cite{Crain1}, Au/Si(5512) \cite{Lee} and Ga/Si(112) 
\cite{Gonzales}. The Au/Si(335) surface consists of Si(111) terraces 
$3\frac{2}{3} \times a_{[11\bar{2}]}$ wide and long monoatomic chains along 
these terraces are observed at Au coverage $0.28 \; ML$ 
\cite{Crain1}-\cite{Crain}. For the Au/Si(557) terraces are $5\frac{2}{3}$ 
wide and critical coverage is $0.2 \; ML$ \cite{Crain1,Crain}.

Many techniques have been developed in surface science in order to investigate 
the electronic and topographic properties of such systems. Those include: angle
resolved photoemission spectroscopy, low and high energy electron diffraction,
x-ray diffraction. Unfortunately they provide averaged data over the sample 
surface and are not appropriate for very small (nanometer size) objects. One of 
the most powerful techniques in surface science is the scanning tunneling 
microscopy (STM) \cite{Hofer} which is extremely sensitive to the electronic 
structure of the topmost atomic layer of the material and allows to study 
properties of the objects on sub-nanometer (atomic) scale. 

On the other hand there is also a couple of theoretical papers regarding STM
technique, mainly in context of single impurity on surface 
\cite{Schiller,Plihal} and the surface itself \cite{Hofer}-\cite{Heike}. 
Recently the problem of one dimensional objects on surfaces has also been 
raised, for example STM imaging of Luttinger liquid \cite{Eggert} or molecular 
wires \cite{Calev}. The surface reconstructions of the gold nanowires on 
Si(557) have been investigated by Sanchez-Portal and Martin within first 
principle density functional theory \cite{Sanchez1}. The local density of 
states of Au chains on NiAl(110) surface has also been studied within the same 
technique \cite{Persson}.

In the present work we provide STM data from Au induced structures on Si(335) 
surface and demonstrate how the STM images depend on the tip-surface voltage, 
especially on its polarization. To understand this effect we propose a model of
tunneling between STM tip and the surface and calculate the tunneling current 
within nonequilibrium Keldysh Green function formalism. Theoretical description 
is in a good qualitative agreement with STM experimental data. Rest of the 
paper is organized as follows: in Sec.\ref{exp} we describe the experimental 
setup and provide some data. In Sec. \ref{model} we introduce theoretical one 
band tight binding model and results are presented in Sec. \ref{results}. 
Finally we conclude in Sec. \ref{concl}.


\section{Experiment \label{exp}}

Experimental setup consists of ultra high vacuum (UHV) chamber equipped with a
scanning tunneling microscope (type OmicronVT) and reflection high energy 
electron diffraction (RHEED) apparatus. Samples were prepared in-situ and the 
base pressure was less than $5 \times 10^{-11}$ mbar during measurements. The 
monoatomic long wires have emerged after deposition of $0.28$ ML of Au and 
heating the sample at temperature $950$ K for $20$ s. The quality of the 
surface reconstruction has been controlled by RHEED technique. 

In Fig. 1 we show the STM topography data of the same area of the
sample taken at tunneling current $I = 0.05$ nA and the sample bias $U = - 1.0$ 
V (a) and $+ 1.0$ V (b).
\begin{figure}[h]
\begin{center}
\includegraphics[width=\textwidth]{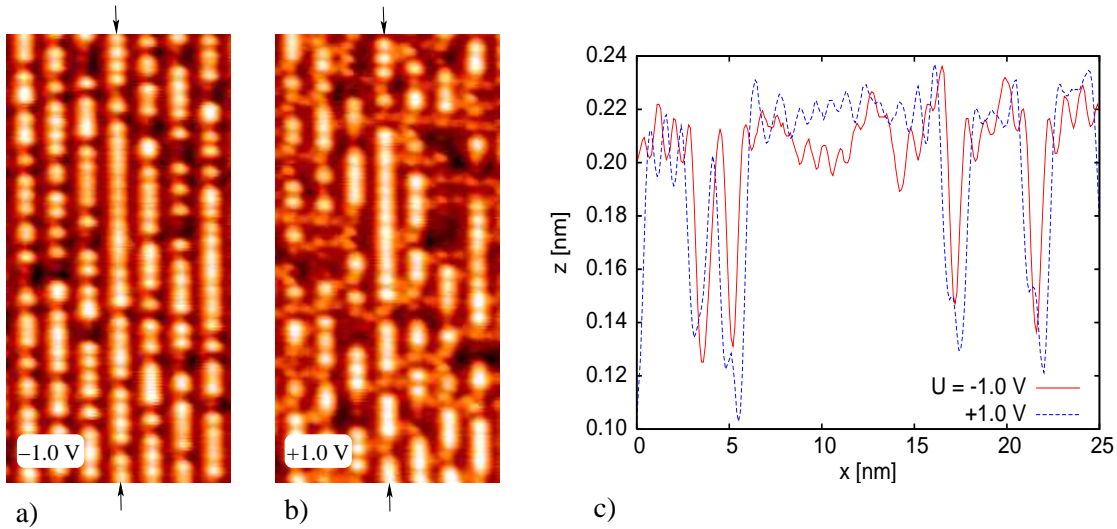}
\caption{The $12.5 \times 25$ nm$^2$ STM topography images of the same area of 
         the Si(335) surface taken at two different sample biases $U = - 1.0$ V 
	 (a) and $U = 1.0$ V (b) with the tunneling current $I = 0.05$ nA. (c) 
	 shows the cross sections along the chain indicated by arrows on (a) 
	 and (b).}
\end{center}
\label{fig1}
\end{figure}
Note clear change of the images. At positive bias some bright structures are no
more visible whereas the others emerge. While the mean features, namely the 
chain - like structure remains, the lengths of the wires and their internal 
structure change a little. It is clearly seen in the Fig. 1 (c), where the 
cross sections along atomic chains, indicated by arrows in (a) and (b), are 
shown. We see that upon reversing of the sample bias reverses also topography 
of the nanowire. The maxima change into minima and vice versa. We stress that 
this effect has strictly electronic origin. In order to avoid possible 
influence of the sample thermal drift, insensitive to the sample bias, the 
specific surface defect (not shown in the figure) was chosen as a reference 
point. 

Those data clearly indicate that the electron density of states along the 
nanowire is strongly modulated with the periodicity, corrected by the
sensitivity of the STM scanner, was equal to about 
$2 \times a_{[1 \bar{1} 0]}$. The reversal of the topography images shows 
that the electron density distribution between the atoms of the nanowire is 
opposite to the electron density distribution on it. To explain this effect we 
have performed calculations within tight binding model.


\section{The model \label{model}}

Our model system is described by the Hamiltonian:
\begin{eqnarray}
H = \sum_{\lambda \in \{t,s\}{\bf k}\sigma} \epsilon_{\lambda {\bf k}} 
c^+_{\lambda {\bf k} \sigma} c_{\lambda {\bf k} \sigma} +
\sum_{\sigma} \varepsilon_d c^+_{d\sigma} c_{d\sigma} +
\sum_{i\sigma} \varepsilon_i c^+_{i\sigma} c_{i\sigma} 
\nonumber \\
+ \sum_{{\bf k}\sigma} \left( t_{td} c^+_{t {\bf k} \sigma} c_{d\sigma} + H. c. 
\right) +
\sum_{i {\bf k}\sigma} \left( t_{is} c^+_{s {\bf k} \sigma} c_{i\sigma} + H. c. 
\right) +
\sum_{\sigma} \left( t_{id} c^+_{d\sigma} c_{i\sigma} + H. c. \right)
\label{hamilt}
\end{eqnarray}
and consists of metallic wire with the site (atomic) energies $\varepsilon_i$ 
and the hopping parameter $t_{ij}$ between atoms. The wire is connected via
parameters $t_{is}$ to the surface, which we treat as a reservoir for 
electrons with single particle energies $\epsilon_{s\bf k}$. Above the wire
there is a tip modeled by single atom with atomic energy $\varepsilon_d$
attached to the another reservoir (with electron energies $\epsilon_{t\bf k}$) 
via hopping $t_{td}$. Tunneling of the electrons between STM tip and one of the 
atoms in a wire is described via tunneling matrix element $t_{id}$. As usually 
$c^+_{\lambda}$ ($c_{\lambda}$) stands for creation (annihilation) electron 
operator in STM lead ($\lambda = t$), tip atom ($\lambda = d$), wire 
($\lambda = i$) and surface ($\lambda = s$). Schematic view of our system is 
shown in the Fig. 2.
\begin{figure}[h]
\begin{center}
\includegraphics[width=0.5\textwidth]{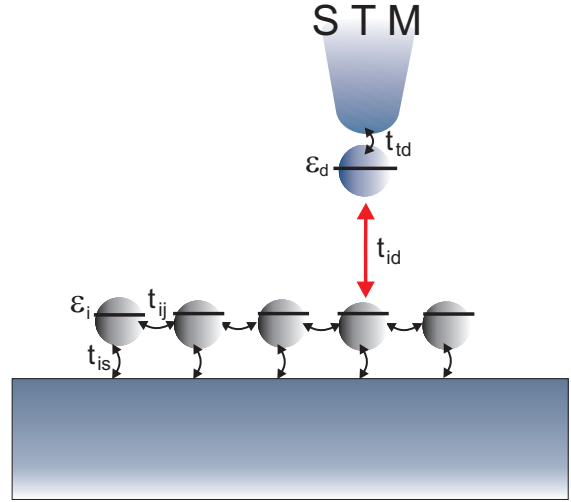}
\caption{Schematic view of the model STM system.}
\end{center}
\label{fig2}
\end{figure}

In order to calculate the tunneling current from the tip to the surface we have
adapted equation of motion technique for nonequilibrium Keldysh Green functions 
\cite{MK1,MK2}. The final expression for the tunneling current through the wire
containing $N$ atoms is given by:
\begin{eqnarray}
I = \frac{2e}{h} \int d\omega {\rm T}(\omega) 
(f(\omega - \mu_t) - f(\omega - \mu_s)) 
\label{current}
\end{eqnarray}
where transmittance is given in the form:
\begin{eqnarray}
{\rm T}(\omega) = \Gamma_t(\omega) \Gamma_s(\omega) 
\sum^N_{i=1} |G_{i d}(\omega)|^2
\label{transmit}
\end{eqnarray}
$\Gamma_{t (s)}(\omega) = 
\Sigma_{\bf k} t_{td (is)} \delta(\omega - \epsilon_{{\bf k} t(s)})$ is the 
coupling parameter between tip lead and tip site, and surface and wire.
$f(\omega)$ is the usual Fermi distribution function and $\mu_t$ ($\mu_s$) is 
the chemical potential of the tip (surface). $G_{id}(\omega)$ is the matrix 
element (connecting the tip site $d$ with the wire atom $i$) of the retarded 
Green function $\hat G(\omega)$, solution of the equation:
\begin{eqnarray} 
\left(\omega \hat 1 - \hat H^{-1} \right) \hat G(\omega) = \hat 1
\label{Green}
\end{eqnarray}
%

 
\section{Results and discussion \label{results}}

To calculate the topography of the wire we have solved Eqs. 
(\ref{current})-(\ref{Green}) selfconsistently for given bias voltage 
($eV = \mu_t - \mu_s$) and fixed current $I$ and obtained value of 
$\Gamma_{id} \equiv |t_{id}|^2$ (in units of $\Gamma_s = \Gamma_t$) which 
depends on the distance between the tip and the wire. In order to make 
comparison with experimental data we have adapted formula from 
Ref. \cite{Calev} connecting these two quantities
\begin{eqnarray}
\Gamma_{id} = \frac{\hbar^3 \pi^2 \sqrt{2 m W_f}}
{m^2 l^3_M \left(\frac{\hbar^2 \pi^2}{2 m l^2_M} + W_f \right)} 
\; e^{- \frac{z}{\hbar} \sqrt{2 m W_f}}
\label{dist}
\end{eqnarray}
where $m$ is the electron mass, $W_f$ - work function taken as $4.5$ eV for 
Si, $l_M$ - length parameter (order of an orbital spatial size) chosen as 
$0.2$ nm \cite{Calev}, and $z$ - distance between tip and the surface. In our 
model the wire atoms have been equally placed with the distance equal to the Si 
lattice constant in direction $[1 \bar{1} 0]$.

In Fig. 3 we show the comparison of the experimental topographic data 
for a wire of typical length observed in STM experiment with the theoretical 
calculations. 
\begin{figure}[h]
\begin{center}
\includegraphics[width=0.49\textwidth]{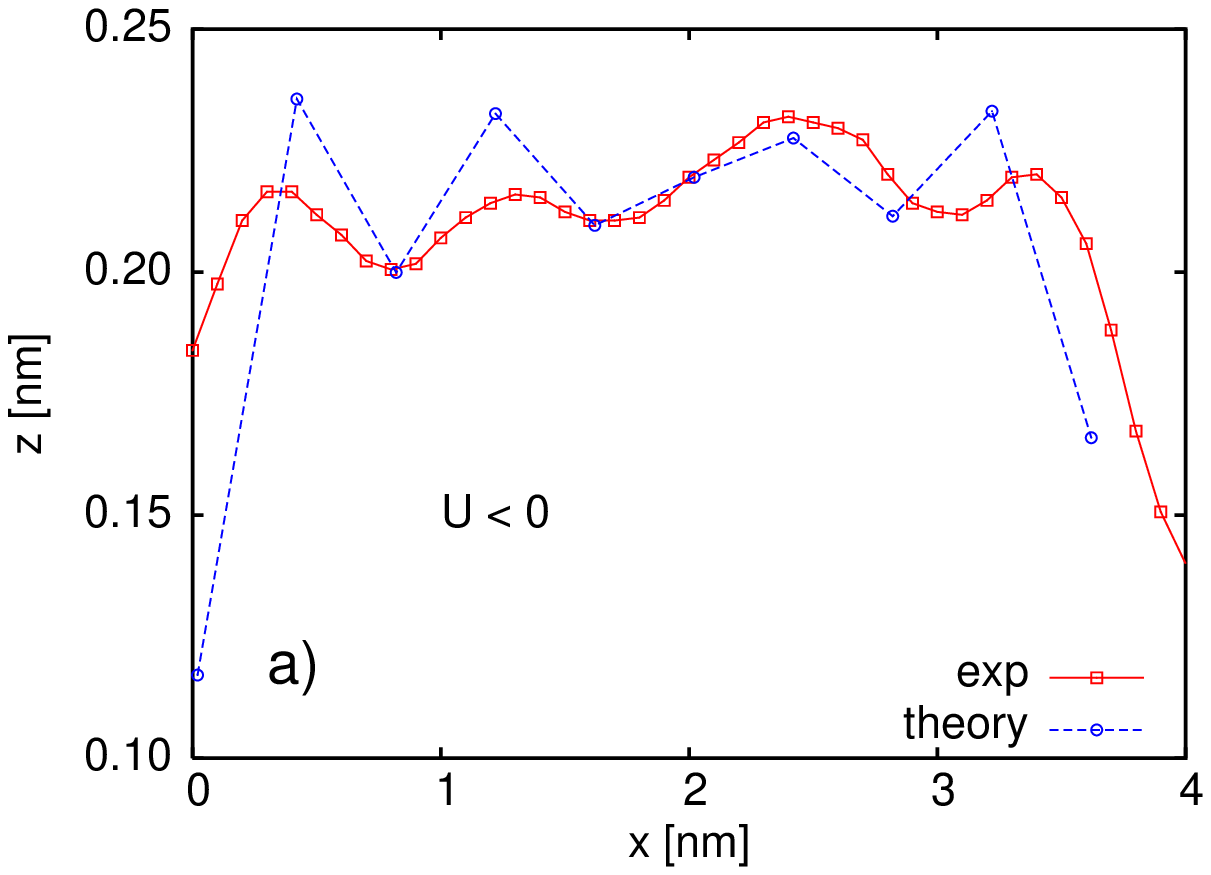}
\includegraphics[width=0.49\textwidth]{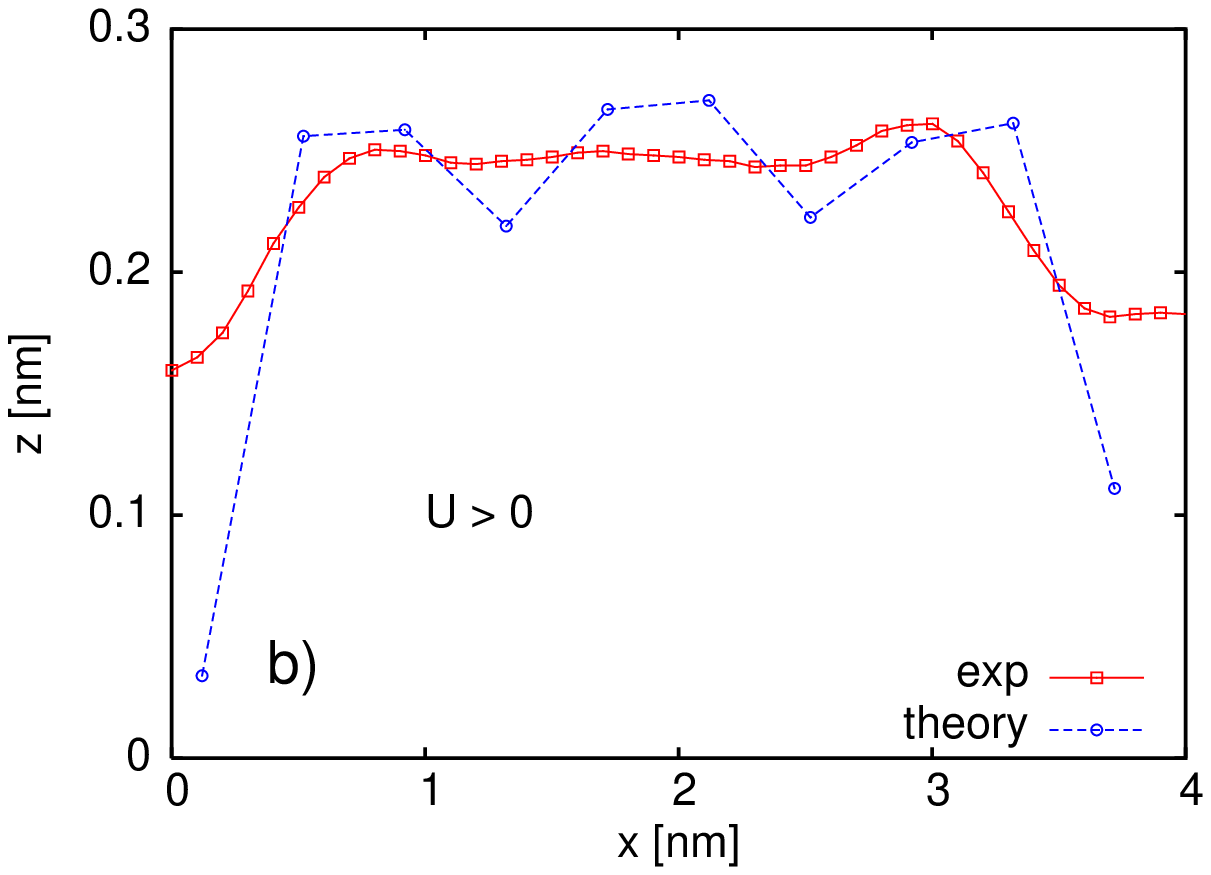}
\caption{Comparison of experimental (squares) and theoretical (circles) data 
         of the distance between STM tip and the surface along one of the
	 typical wires for negative (a) and positive (b) sample bias. For 
	 details see text.}
\end{center}
\label{fig3}
\end{figure}

First of all it is worthwhile to note that we observe different number of 
minima and maxima of the distance between tip and surface for negative 
Fig. 3 (a) and positive (b) bias voltages. While for $U < 0$ we have 
four maxima, for $U > 0$ there are 3 such maxima. This is due to different 
energy distribution of the transmittance ${\rm T}(\omega)$ 
(see Eq. (\ref{transmit})). The transmittance ${\rm T}(\omega)$  as a function
of energy $\omega$ for positive and negative sample bias is shown in the 
Fig. 4.  
\begin{figure}[h]
\begin{center}
\includegraphics[width=0.5\textwidth]{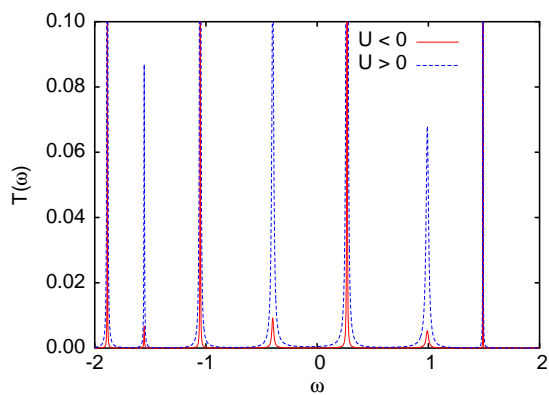}
\caption{Transmittance (Eq. \ref{transmit}) as a function of energy $\omega$ 
         for positive and negative sample bias. The STM tip position is above 
	 an atom in the middle of the wire. Parameters are the same as in 
	 Fig. 3.}
\end{center}
\label{fig4}
\end{figure}
For the energy independent $\Gamma_{t(d)}$ it is in close relation with the 
wire density of states. The transmittances integrated over energy together 
with the Fermi functions give tunneling currents and therefore different values 
of the distance which is in close relation with $\Gamma_{id}$ via Eq. 
(\ref{dist}). So one can say that this effect is purely of electronic
nature. Similar effect, namely changes of the number of maxima and minima 
in the local density of states with bias voltage, has been observed in 
Ref. \cite{Persson} for gold wires on NiAl(110) surface.

Another important effect is the period of the oscillations seen in the 
Fig. 3. We stress that the periodicity of the atomic chain, as expected from the
Au coverage \cite{Jalochowski,Zdyb}, is equal to $a_{[1 \bar{1} 0]}$, whilest
the observed and calculated period is equal to about 
$2 \times a_{[1 \bar{1} 0]}$. This period seems to not depend on the wire 
length. Moreover, to get a reasonable consistency with the experimental data, 
we had to modify an atomic energy of the atom at one of the ends of the wire. 
This indicates that wires strongly interact via surface with neighboring ones 
on the same terrace.


\section{Conclusions \label{concl}}

In conclusions we have studied experimentally and theoretically gold nanowires
on vicinal Si(335) surface. The wires of differnent length were grown along 
terraces in direction $[1 \bar 1 0]$ at Au coverage equal to $0.28$ ML. The STM 
topographic data show strong dependence on the sample bias revealing space 
dependent oscillations of the distance between STM tip and surface with a 
period two of Si lattice constant in direction $[1 \bar 1 0]$, along the chain. 
The period of oscillations seems to not depend on the length of the nanowires. 
Moreover calculations based on tight binding model indicate that those wires 
strongly interact with neighboring ones situated on the same terrace. The 
problem of interaction with the surface as well as interaction between chains 
on differnt terraces is under consideration and will be published elsewhere.


\begin{acknowledgement}
M. K. acknowledges a partial support by the grant no. PBZ-MIN-008/P03/2003. 
\end{acknowledgement}


\end{document}